\documentclass[twocolumn,showpacs,preprintnumbers,amsmath,amssymb,nofootinbib]{revtex4-1}
\usepackage{CJK}

\usepackage{graphicx}
\usepackage{dcolumn}
\usepackage{bm}
\usepackage{amsfonts}
\usepackage{mathrsfs}
\usepackage{verbatim}

\def\bequ{\begin{equation}}
\def\eequ{\end{equation}}
\def\be{\begin{equation}}
\def\ee{\end{equation}}

\begin{document}


\title{Bifurcation of the Maxwell quasinormal spectrum on asymptotically anti--de Sitter black holes}

\author{Mengjie Wang}
\email{Corresponding author: mjwang@hunnu.edu.cn.}
\author{Zhou Chen}
\author{Xin Tong}
\author{Qiyuan Pan}
\email{Corresponding author: panqiyuan@hunnu.edu.cn}
\author{Jiliang Jing}
\email{jljing@hunnu.edu.cn}
\affiliation{\vspace{2mm}
Department of Physics and Synergetic Innovation Center for Quantum Effects and Applications, Hunan Normal University,
Changsha, Hunan 410081, P.R. China \vspace{1mm}}
\date{\today}

\begin{abstract}
We study the Maxwell quasinormal spectrum on asymptotically anti--de Sitter black holes with a set of two Robin type boundary conditions, by requiring the energy flux to vanish at asymptotic infinity. Focusing, for illustrative purposes, on Schwarzschild--anti--de--Sitter black holes both without and with a global monopole, we unveil that, on the one hand, the Maxwell quasinormal spectrum \textit{bifurcates} as the black hole radius increases for both boundary conditions, and which is termed as the \textit{mode split} effect; while on the other hand, with an appropriate fixed black hole radius but increasing the monopole parameter, the first (second) boundary condition may trigger (terminate) the mode split effect.       
\end{abstract}

\maketitle

\section{Introduction}
Black holes (BHs) are predicted by general relativity, as solutions of the elegant Einstein field equations, characterized by an event horizon at the boundary and a singularity at the origin~\cite{FrolovNovikov}. When a cosmological constant is included in the Einstein equations, one may obtain BH solutions with various asymptotics. Among them, asymptotically anti--de Sitter (AdS) BHs have attracted a lot of attention recently. One reason is that the AdS/CFT correspondence~\cite{Maldacena:1997re,Gubser:1998bc,Witten:1998qj}, which is a correspondence between a gravitation theory and a quantum field theory, may be applied to strongly coupled systems, ranging from condensed matter physics~\cite{Hartnoll:2009sz,Horowitz:2010gk,Cai:2015cya} to nuclear physics~\cite{Mateos:2007ay,Gubser:2009md,CasalderreySolana:2011us}. Another reason is that the timelike property of the AdS boundary may lead to interesting features, as compared to other asymptotic spacetimes, such as the weak turbulent instability~\cite{Bizon:2011gg}, the superradiant instability of massless fields~\cite{Uchikata:2009zz,Cardoso:2013pza,Wang:2015fgp} and the generic Robin type vanishing energy flux boundary conditions~\cite{Wang:2015goa,Wang:2015fgp,Wang:2016dek,Wang:2017fie,Wang:2019qja,Wang:2016zci}. In this paper, we unveil another novel feature of asymptotically AdS BHs, that is \textit{bifurcation} of the Maxwell quasinormal spectrum. 

Black hole quasinormal modes (QNMs), describing the characteristic oscillations of BHs, are defined as eigenvalues of perturbation equations with appropriate boundary conditions~\cite{Kokkotas:1999bd,Berti:2009kk,Konoplya:2011qq}. On asymptotically AdS BHs, one may look for QNMs of various spin fields by imposing an ingoing wave boundary condition at the horizon and at infinity, requiring either Dirichlet type field vanishing boundary conditions~\cite{Horowitz:1999jd,Chan:1996yk,Chan:1999sc,Cardoso:2001bb,Cardoso:2003cj,Berti:2003ud,Giammatteo:2004wp,Jing:2005ux,Jing:2005uy} or Robin type vanishing energy flux boundary conditions~\cite{Wang:2015goa,Wang:2015fgp,Wang:2016dek,Wang:2017fie,Wang:2019qja,Wang:2016zci}. The advantage of the latter Robin type boundary conditions is that they are applicable, on spherically symmetric backgrounds, not only in the Regge--Wheeler--Zerilli~\cite{Regge:1957td, Zerilli:1970se} but also in the Teukolsky formalisms~\cite{Teukolsky:1972my,Teukolsky:1973ha}. In this paper, we focus on the Regge--Wheeler--Zerilli formalism and employ the Robin type vanishing energy flux boundary conditions.

In the study of QNMs on asymptotically AdS BHs, the Maxwell spectrum has a peculiar structure, as compared to other spin fields, in the sense that the real part of some Maxwell modes asymptotes to \textit{zero} for large BHs~\cite{Cardoso:2001bb,Cardoso:2003cj,Berti:2003ud,Wang:2015goa}. To further investigate and understand this behavior, we implement a study in this paper to solve the Maxwell QNMs on Schwarzschild--AdS BHs both without and with a global monopole, by varying the BH radius. We confirm that there is a critical BH radius, associated to each fixed angular momentum quantum number $\ell$ and overtone number $N$, and the real part of the Maxwell QNMs becomes \textit{zero} when the BH size is greater than the critical BH radius. More surprisingly, we further unveil that when the real part of the Maxwell spectrum becomes zero, the imaginary part \textit{branches off}. This is dubbed as the \textit{mode split} effect, and it may be related with spectrum bifurcation observed for near extremal Kerr BHs~\cite{Yang:2012pj}. 

Our results are obtained numerically, by using a pseudospectral method~\cite{trefethen2000spectral}. This method is a robust numeric approach, and has been applied widely in BH physics, ranging from numerical relativity~\cite{Grandclement:2007sb} to BH perturbation theory~\cite{Dias:2009iu}. It is particular efficient to solve QNMs by using the pseudospectral method since an initial guess value is not necessary, and it has been successfully employed to obtain a novel set of purely imaginary modes of scalar fields on Schwarzschild--de Sitter BHs~\cite{Jansen:2017oag}.  

This paper is organized as follows. In Section~\ref{seceq} we introduce the background geometry, i.e. Schwarzschild--AdS BHs both without and with a global monopole, and derive the Maxwell equations and the corresponding boundary conditions on the aforementioned background, in the Regge--Wheeler--Zerilli formalism. In Section~\ref{secnum} we first introduce a numeric pseudospectral method, and then apply it to look for the Maxwell quasinormal spectrum and, in particular, demonstrate the \textit{mode split} effect. Final remarks and conclusions are presented in Section~\ref{discussion}.

\section{background geometry, Maxwell equations and boundary conditions}
\label{seceq}
In this section, we briefly review the background geometry of Schwarzschild--AdS BHs both without and with a global monopole, present equations of motion for the Maxwell fields in the Regge--Wheeler--Zerilli formalism and derive the corresponding boundary conditions.

\subsection{The line element}
We start by considering the following spherically symmetric line element~\cite{Barriola:1989hx,Secuk:2019njc}
\begin{equation}
ds^2=\dfrac{\Delta_r}{r^2}dt^2-\dfrac{r^2}{\Delta_r}dr^2-r^2\left(d\theta^2+\sin^2\theta d\varphi^2\right) \;,\label{metric}
\end{equation}
with the metric function
\begin{equation}
\Delta_r\equiv r^2\left(\tilde{\eta}^2+\frac{r^2}{L^2}\right)-2Mr\;,\label{metricfunc}
\end{equation}
where $L$ is the AdS radius, $M$ is the mass parameter, and the event horizon $r_+$ is determined by the non--zero real root of $\Delta_r(r_+)=0$. Here the dimensionless parameter $\tilde{\eta}^2$ is defined by $\tilde{\eta}^2\equiv 1-8\pi\eta^2$, where $\eta$ represents the global monopole and its effects have been explored in various contexts, see $e. g.$~\cite{Secuk:2019njc,Yu:2002st,Pan:2008xz,Chen:2009vz,Piedra:2019ytw}. In the limit of $\eta^2=0$, Schwarzschild--AdS spacetimes may be recovered.

The asymptotic structure of the spacetime, given by Eq.~\eqref{metric}, can be obtained by taking the limit $r\rightarrow\infty$ 
\begin{align}
ds^2=&\left(1+\frac{\tilde{r}^2}{L^2}\right)d\tilde{t}^{2}-\left(1+\frac{\tilde{r}^2}{L^2}\right)^{-1}d\tilde{r}^2\nonumber\\&-\tilde{\eta}^2\tilde{r}^2(d\theta^2+\sin^2\theta d\varphi^2)\;,\label{asymetric}
\end{align}
where $\tilde{t}=\tilde{\eta}t\;,\tilde{r}=\tfrac{r}{\tilde{\eta}}$
and which describes a pure AdS spacetime with a solid deficit angle $4\pi\tilde{\eta}^2$. 

\subsection{Equations of motion in the Regge--Wheeler--Zerilli formalism}
In a spherically symmetric background, one may obtain variable separated and degrees of freedom decoupled Maxwell equations in the Regge--Wheeler--Zerilli formalism~\cite{Regge:1957td, Zerilli:1970se}.

We start from the Maxwell equations
\begin{equation}
\nabla_{\nu}F^{\mu\nu}=0\;,\label{Maxwelleq}
\end{equation}
where the field strength tensor is defined as $F_{\mu\nu}=\partial_{\mu}A_{\nu}-\partial_{\nu}A_{\mu}$. 
We then expand the vector potential $A_\mu$ in terms of the scalar and vector spherical harmonics~\cite{Ruffini:1973}
\begin{equation}
A_{\mu}=e^{-i\omega t}\sum_{\ell, m}\left(\left[\begin{array}{c} 0 \\
0 \\
a^{\ell m}(r) \boldsymbol {S}_{\ell m}\end{array}\right]+\left[
\begin{array}{c}j^{\ell m}(r)Y_{\ell m}  \\
h^{\ell m}(r)Y_{\ell m} \\
k^{\ell m}(r)\boldsymbol {Y}_{\ell m}  
\end{array}\right]\right)\;,\label{Vpotential}
\end{equation}
with the definition of the vector spherical harmonics
\begin{equation}
\boldsymbol {S}_{\ell m}=
\left(\begin{array}{c} \frac{1}{\sin \theta} \partial_{\varphi}Y_{\ell m}  \\
-\sin \theta \partial_{\theta}Y_{\ell m}\end{array}\right)\;,\;\;\;
\boldsymbol {Y}_{\ell m}=
\left(\begin{array}{c} \partial_{\theta}Y_{\ell m}  \\
\partial_{\varphi}Y_{\ell m}\end{array}\right)\;,\nonumber
\end{equation}
where $Y_{\ell m}$ are the scalar spherical harmonics, $\omega$ is the frequency, $m$ is the azimuthal number, $\ell$ is the angular momentum quantum number. Note that the first term in the right hand side of Eq.~\eqref{Vpotential} has parity $(-1)^{\ell+1}$ and the second term has parity $(-1)^\ell$, and we shall call the former (latter) the axial (polar) modes. By substituting Eq.~\eqref{Vpotential} into Eq.~\eqref{Maxwelleq}, one obtains the Schrodinger-like radial wave equation 
\begin{equation}
\left(\frac{d^2}{dr_{*}^2}+\omega^2-\ell(\ell+1)\dfrac{\Delta_r}{r^4}\right)\Psi(r)=0\;,\label{RWZeq}
\end{equation}
where the tortoise coordinate is defined as 
\begin{equation}
\dfrac{dr_*}{dr}=\dfrac{r^2}{\Delta_r}\;,\label{tortoisecoor}
\end{equation}
with $\Psi(r)=a^{\ell m}(r)$ for axial modes, and 
\begin{equation}
\Psi(r)=\dfrac{r^2}{\ell(\ell+1)}\left(-i\omega h^{\ell m}(r)-\dfrac{dj^{\ell m}(r)}{dr}\right)\;,\nonumber
\end{equation}
for polar modes.
\subsection{Boundary conditions}
In order to solve the radial equation~\eqref{RWZeq}, we impose an ingoing wave boundary condition at the horizon, as usual. At infinity, we require that energy flux should be vanished at asymptotic infinity, following a generic principle we proposed in~\cite{Wang:2015goa} (see also~\cite{Wang:2016dek,Wang:2016zci}).

We start from the energy--momentum tensor of the Maxwell field, which is given by
\begin{equation}
T_{\mu \nu}=F_{\mu\sigma}F^\sigma_{\;\;\;\nu}+\dfrac{1}{4}g_{\mu\nu}F^2\;.\label{EMTensor}
\end{equation}
Then the spatial part of the radial energy flux may be calculated as 
\begin{equation}
\mathcal{F}|_r\propto\dfrac{\Delta_r}{r^2}\Psi(r)\Psi^\prime(r)\;,\label{RWZbc1}
\end{equation}
where $\prime$ denotes a derivative with respect to $r$.
By expanding Eq.~\eqref{RWZeq} asymptotically
\begin{equation}
\Psi\sim c_{0}+\frac{c_{1}}{r}+\mathcal{O}\left(\frac{1}{r^2}\right)\;,\label{RWZasysol}
\end{equation}
Eq.~\eqref{RWZbc1} becomes $\mathcal{F}|_{r,\infty}\propto c_0c_1$.
\\
Then the vanishing energy flux principle, i.e. $\mathcal{F}|_{r,\infty}=0$, leads to the following two solutions 
\begin{align}
&c_0=0\;,\label{RWZbc2-1}\\
&c_1=0\;.\label{RWZbc2-2}
\end{align}

\section{Numerics}
\label{secnum}
In order to study the Maxwell QNMs thoroughly in a full parameter space, numeric methods are resorted. In this part, we first introduce a numeric pseudospectral method~\cite{trefethen2000spectral}, and then present a few selected results obtained by this method and, in particular, illustrate the \textit{mode split} effect.

\subsection{Method}
For numeric convenience, we first transform Eq.~\eqref{RWZeq} from a quadratic eigenvalue problem into a linear eigenvalue problem by
\begin{equation}
\Psi=e^{-i\omega r_\ast}\phi\;,\label{spectraltrans}
\end{equation}
where the tortoise coordinate $r_\ast$ is defined in Eq.~\eqref{tortoisecoor}. Then changing the coordinate from $r$ to $z$ through
\begin{equation}
z=1-\dfrac{2r_+}{r}\;,\label{rtoz}
\end{equation}
which brings the integration domain from $r\in[r_+,\infty]$ to $z\in[-1,+1]$, Eq.~\eqref{RWZeq} turns into the following form
\begin{equation}
\mathcal{B}_0(z)\phi(z)+\mathcal{B}_1(z,\omega)\phi^\prime(z)+\mathcal{B}_2(z)\phi^{\prime\prime}(z)=0\;,\label{spectraleq1}
\end{equation}
where $\prime$ denotes a derivative with respect to $z$. Here each of the functions $\mathcal{B}_j (j=0,1,2)$ can be derived straightforwardly by substituting Eqs.~\eqref{spectraltrans} and~\eqref{rtoz} into Eq.~\eqref{RWZeq}, and $\mathcal{B}_1$ is linear in $\omega$, i.e. $\mathcal{B}_1(z,\omega)=\mathcal{B}_{1,0}(z)+\omega\mathcal{B}_{1,1}(z)$.

The pseudospectral method solves a differential equation by replacing a continuous variable with a set of discrete grid points. For that purpose, we introduce the Chebyshev points
\begin{equation}
z_j=\cos\left(\dfrac{j\pi}{n}\right)\;,\;\;\;\;\;\;j=0,1,...,n\;,\label{spectralpoints}
\end{equation}
where $n$ denotes the number of grid points. One may construct the Chebyshev differentiation matrices $D^{(1)}$ by using these points~\cite{trefethen2000spectral} and apply them to differentiate $\phi(z)$. Then Eq.~\eqref{spectraleq1} becomes a standard eigenvalue problem
\begin{equation}
(M_0+\omega M_1)\phi(z)=0\;,\label{spectraleq2}
\end{equation}
where $M_0$ and $M_1$ are matrices, $(M_0)_{ij}=\mathcal{B}_{0}(z_i)\delta_{ij}+\mathcal{B}_{1,0}(z_i)D^{(1)}_{ij}+\mathcal{B}_{2}(z_i)D^{(2)}_{ij}$, and similarly for $M_1$. For simplicity, we define the second order Chebyshev differential matrix $D^{(2)}$ by squaring the first order Chebyshev differential matrix $D^{(1)}$~\cite{trefethen2000spectral}.

To solve the eigenvalue equation~\eqref{spectraleq2}, we impose a regular boundary condition at the horizon, since from Eq.~\eqref{spectraltrans} an ingoing wave boundary condition is satisfied automatically for $\phi$. At infinity, from Eqs.~\eqref{spectraltrans} and~\eqref{RWZasysol}, one obtains
\begin{equation}
\phi=0\;,\;\;\;\;\;\;\dfrac{\phi^\prime}{\phi}=\dfrac{i\omega L^2}{2r_+}\;,\label{bcnumerical}
\end{equation}
corresponding to the boundary conditions given by Eq.~\eqref{RWZbc2-1} and Eq.~\eqref{RWZbc2-2}, respectively, and where $'$ again denotes a derivative with respect to $z$.

\subsection{Results}
In the numerical calculations all physical quantities are measured by the AdS radius $L$ so that we take $L=1$. The results presented below are computed by the pseudospectral method described in the above subsection, and they are double checked by a direct integration and the Horowitz--Hubeny approaches adapted from our previous works~\cite{Herdeiro:2011uu,Wang:2012tk,Wang:2014eha,Wang:2017fie,Wang:2019qja}. Note that we use $\omega_1$ ($\omega_2$) to represent QNMs corresponding to the first (second) boundary condition, given by Eq.~\eqref{RWZbc2-1} (Eq.~\eqref{RWZbc2-2}). Moreover, $\ell$ and $N$ are introduced to denote the angular momentum quantum number and the overtone number.
\subsubsection{Without a global monopole}
\begin{figure*}
\begin{center}
\begin{tabular}{c}
\hspace{-4mm}\includegraphics[clip=true,width=0.326\textwidth]{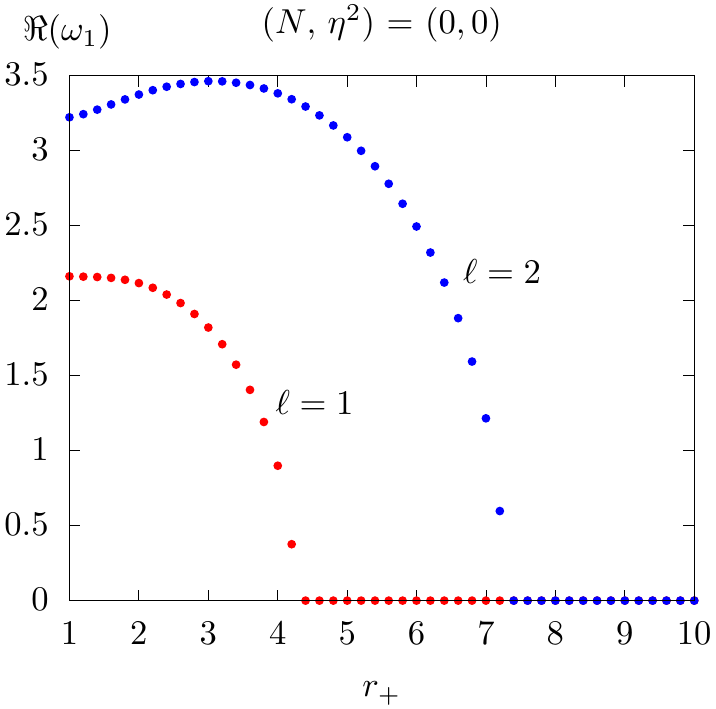}\hspace{12mm}\includegraphics[clip=true,width=0.326\textwidth]{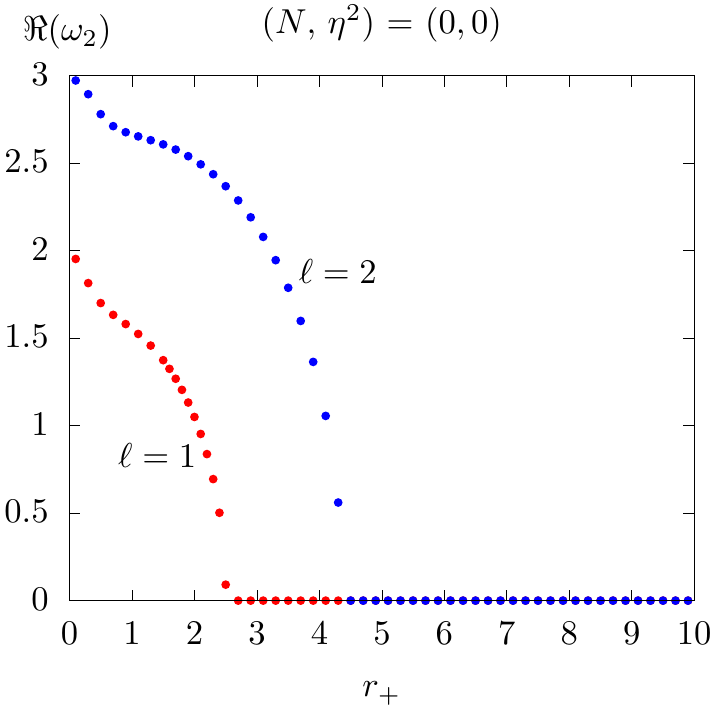}
\\
\hspace{-2mm}\includegraphics[clip=true,width=0.322\textwidth]{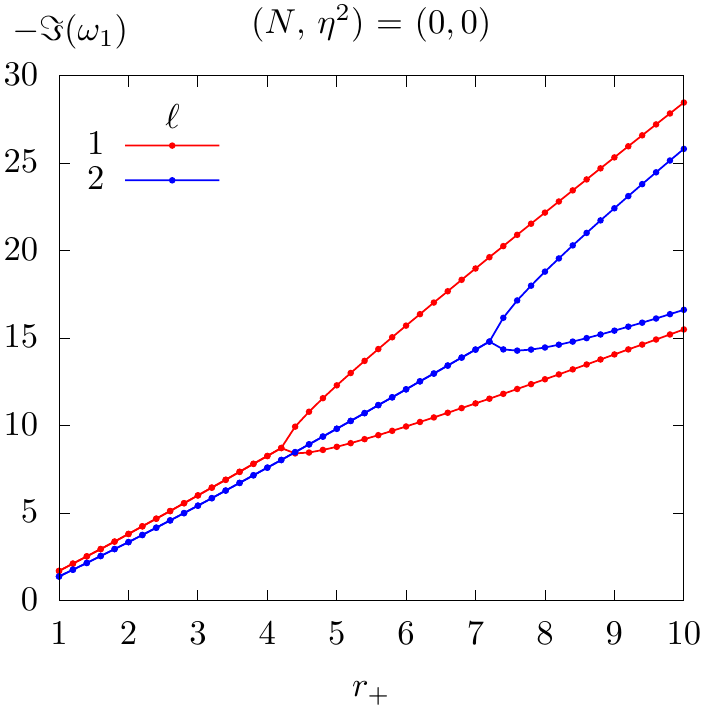}\hspace{13.5mm}\includegraphics[clip=true,width=0.322\textwidth]{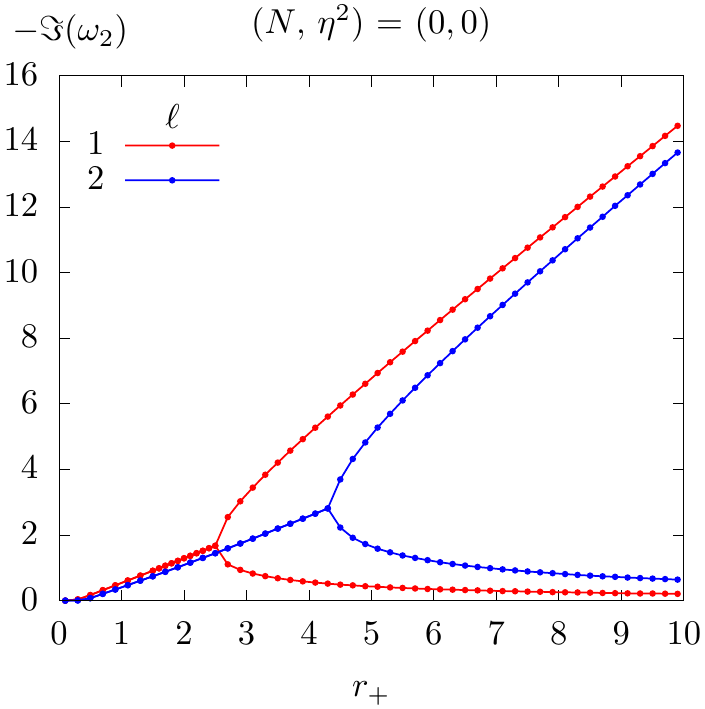}
\end{tabular}
\end{center}
\caption{\label{Fig_splitN0} Real (top) and imaginary (bottom) parts of the Maxwell QNMs on Schwarzschild--AdS BHs vs BH size $r_+$  for $\ell=1, 2$ and $N=0$, with the first (left) and second (right) boundary conditions. The critical BH radius with the first (second) boundary condition is $r^c_{+1}\approx4.24057$ ($r_{+2}^c\approx2.50330$) for $\ell=1$ and is $r^c_{+1}\approx7.26167$ ($r_{+2}^c\approx4.37526$) for $\ell=2$. Precisely at the critical BH radius, the real part of QNMs vanishes and the imaginary part splits into two branches.}
\end{figure*}

We first present both real and imaginary parts of the Maxwell QNMs in terms of the BH size $r_+$ in Fig.~\ref{Fig_splitN0}, by taking $\ell=1, 2$ and $N=0$ as examples, and for both boundary conditions.

A striking feature one may observe is that when the BH size $r_+$ is larger than the critical BH radius $r^c_+$, the real part of QNMs turns into \textit{zero} while the imaginary part \textit{branches off} into two sets of modes. This phenomenon which, as far as we know, has never been reported in the literature and is thus dubbed as the \textit{mode split} effect. 

More precisely, it is shown clearly in Fig.~\ref{Fig_splitN0} that, for both boundary conditions, when the BH size is smaller than the critical BH radius, the magnitude of the imaginary part of QNMs increases as $r_+$ increases and it scales almost linearly with $r_+$. While when the BH size is larger than the critical BH radius, since the mode branches off, we dub, in magnitude, the larger one as the \textit{upper mode} and the smaller one as the \textit {lower mode}. The upper mode, for both boundary conditions, always increases as $r_+$ increases and it scales linearly with $r_+$ when the BH size is away from the critical BH radius. The lower mode, on the other hand, first decreases and then increases (decreases) as $r_+$ increases and it scales linearly with $r_+$ (scales with $r_+$ as $1/r_+$) when the BH size is away from the critical BH radius, for the first (second) boundary condition.

\begin{figure*}
\begin{center}
\begin{tabular}{c}
\includegraphics[clip=true,width=0.323\textwidth]{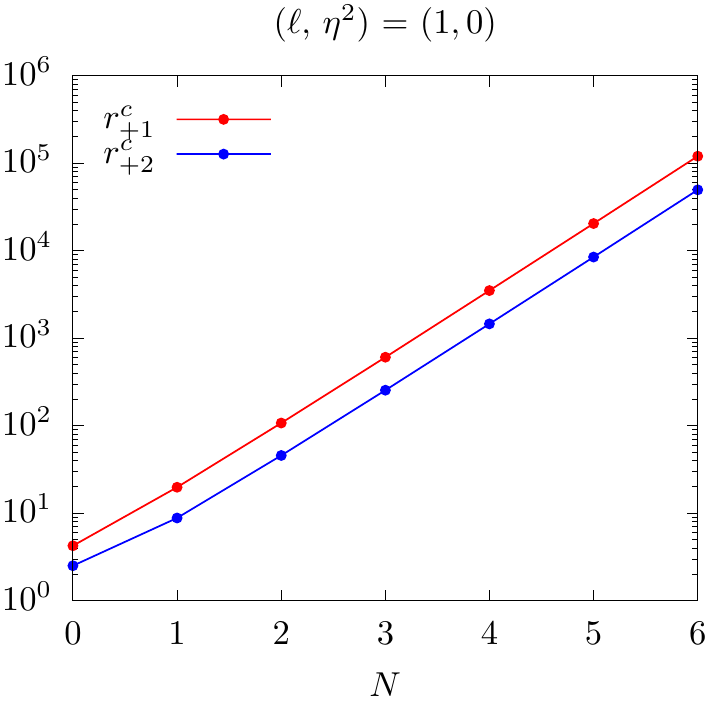}\hspace{12mm}
\hspace{2mm}\includegraphics[clip=true,width=0.323\textwidth]{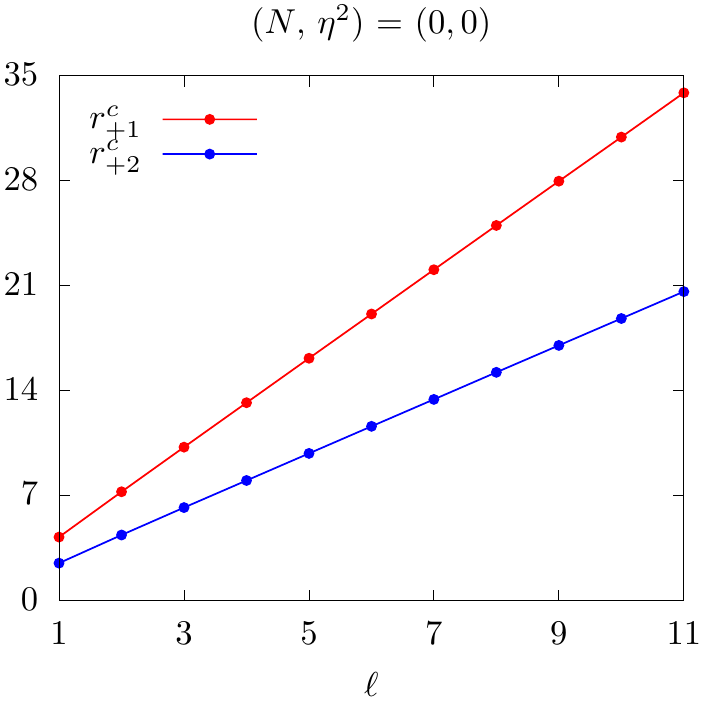}
\end{tabular}
\end{center}
\caption{\label{Fig_criticalradius} The critical BH radius, denoted by $r_{+1}^c$ ($r_{+2}^c$) for the first (second) boundary condition, versus the overtone number $N$ (left) with fixed $\ell=1$ and the angular momentum quantum number $\ell$ (right) with fixed $N=0$. Note that in the left panel, we use the logarithmic scale in the vertical axis.}
\end{figure*}

The mode split effect \textit{does} exist for various values of $\ell$ and $N$. Since the qualitative behavior of QNMs, by fixing $\ell$ and $N$ as other values, is quite similar to the cases shown in Fig.~\ref{Fig_splitN0}, we instead show the critical BH radius by varying $N$ and $\ell$ in Fig.~\ref{Fig_criticalradius}. In the left panel, we present the critical BH radius, denoted by $r_{+1}^c$ ($r_{+2}^c$) for the first (second) boundary condition, by varying $N$ in the semilogarithmic coordinate, and it indicates that critical BH radius for excited modes grows exponentially with $N$, i.e. in the semilogarithmic coordinate scales linearly with $N$, and shares the same slope for both boundary conditions. In the right panel, we display the critical BH radius with $\ell$, and it indicates clearly that the critical BH radius scales linearly with $\ell$ for both boundary conditions but grows faster with the first boundary condition.

\subsubsection{With a global monopole}
\begin{figure*}
\begin{center}
\begin{tabular}{c}
\includegraphics[clip=true,width=0.323\textwidth]{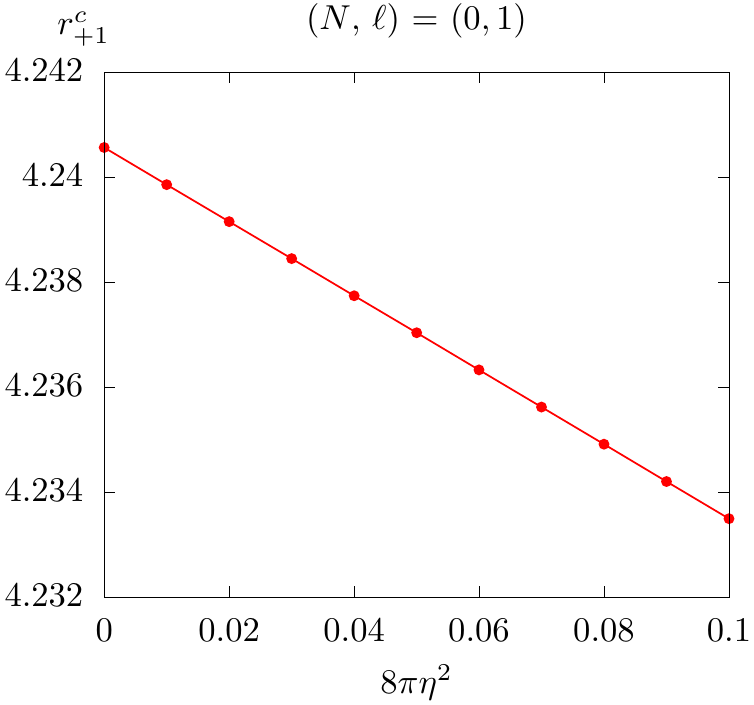}\hspace{11mm}
\includegraphics[clip=true,width=0.320\textwidth]{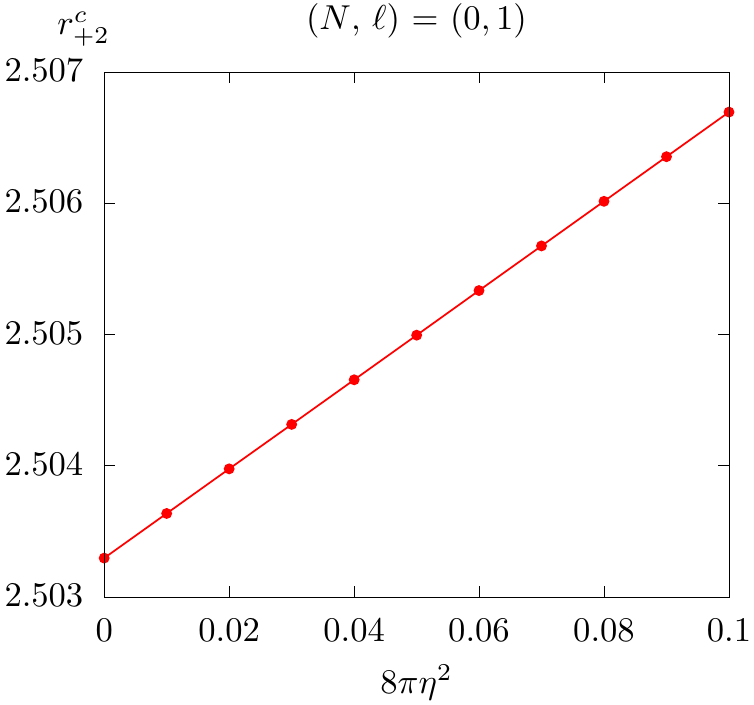}
\end{tabular}
\end{center}
\caption{\label{Fig_criticalmonopole} The critical BH radius ($r_+^c$) versus the monopole parameter ($8\pi\eta^2$), with fixed $\ell=1$ and $N=0$ for the first (left) and second (right) boundary conditions.}
\end{figure*}

By turning on and fixing the monopole parameter $8\pi\eta^2$, the mode split effect still holds. To further analyze this effect for the monopole case, we calculate the critical BH radius, with fixed $N=0$ and $\ell=1$, by varying the monopole parameter $8\pi\eta^2$, and the corresponding results are presented in Fig.~\ref{Fig_criticalmonopole}. As one may observe, the critical BH radius with the first (second) boundary condition decreases (increases) as $8\pi\eta^2$ increases, and for both cases the critical BH radius scales linearly with $8\pi\eta^2$. 

\begin{figure*}
\begin{center}
\begin{tabular}{c}
\hspace{4mm}\includegraphics[clip=true,width=0.322\textwidth]{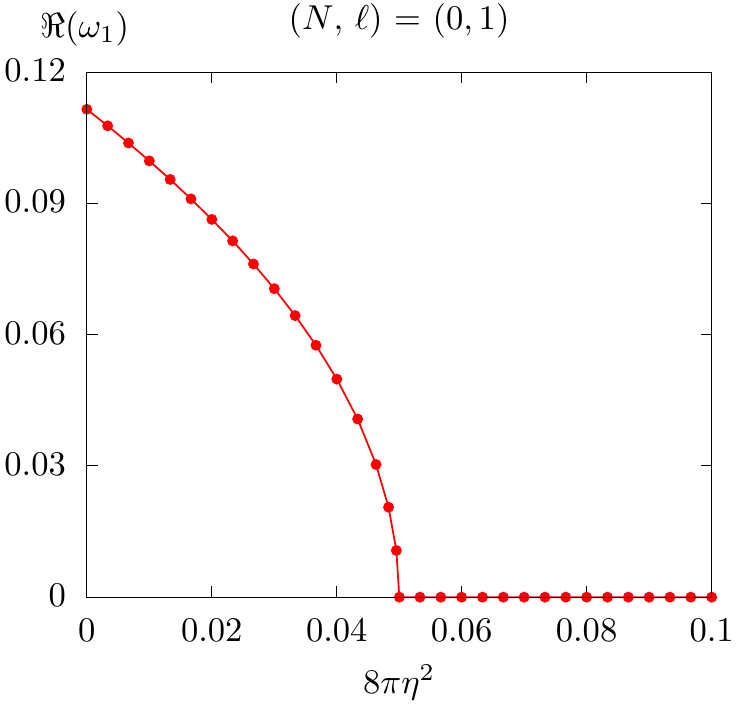}\hspace{12mm}\includegraphics[clip=true,width=0.320\textwidth]{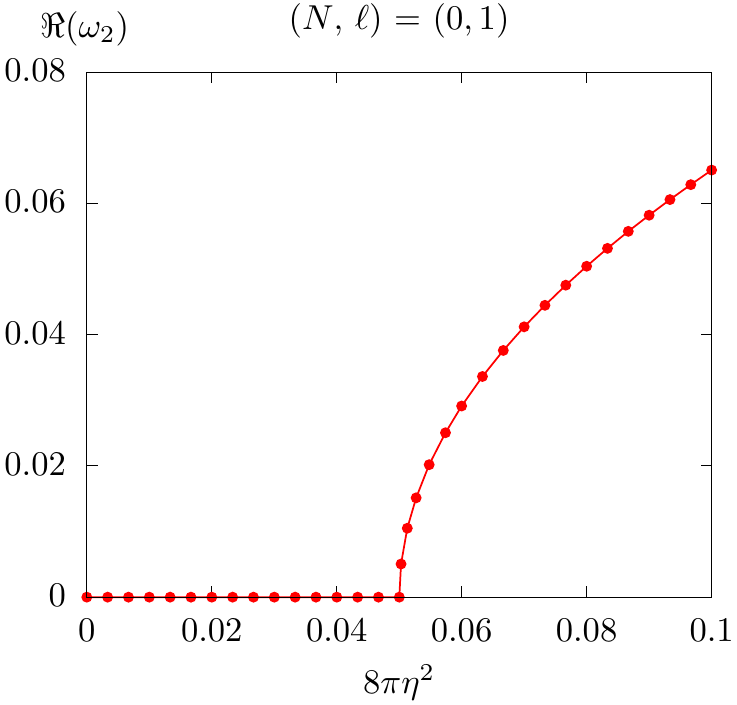}
\\
\hspace{6mm}\includegraphics[clip=true,width=0.326\textwidth]{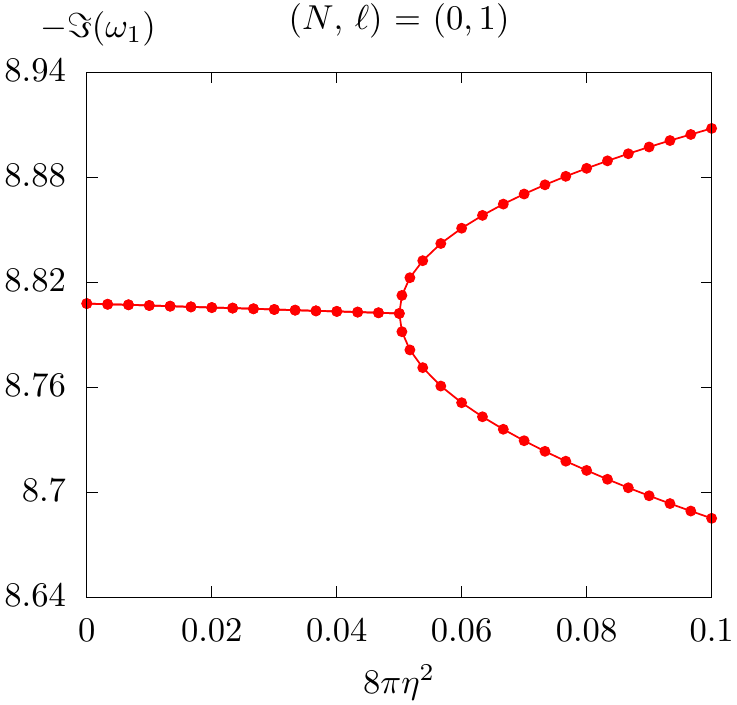}\hspace{12mm}\includegraphics[clip=true,width=0.323\textwidth]{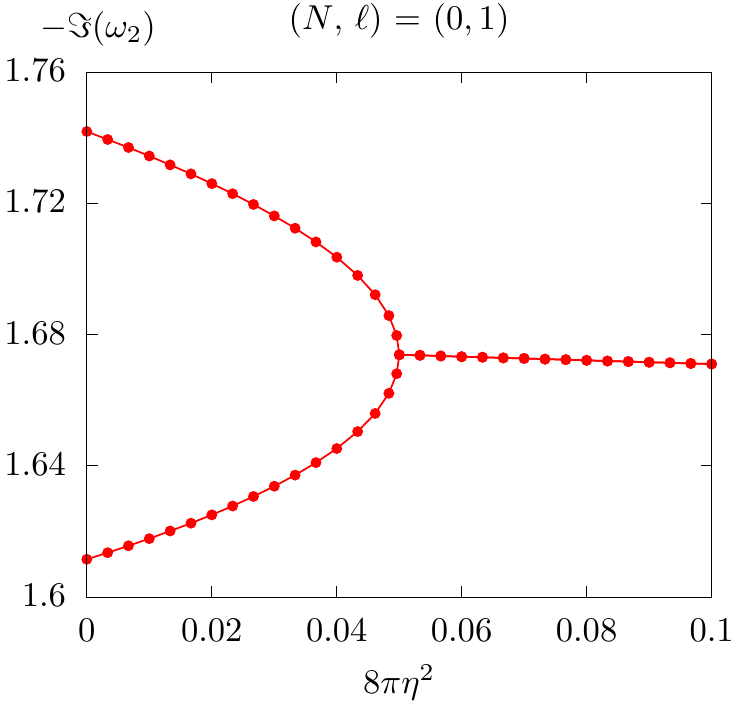}
\end{tabular}
\end{center}
\caption{\label{Fig_monopole} Real (top) and imaginary (bottom) parts of QNMs for Maxwell fields on global monopole--Schwarzschild--AdS BHs vs the monopole parameter $8\pi\eta^2$, with fixed $\ell=1$, $N=0$, $r_{+1}\approx4.237$ for the first (left) and $r_{+2}\approx2.505$ for the second (right) boundary conditions.}
\end{figure*}

The property shown in Fig.~\ref{Fig_criticalmonopole} leads directly to a consequence that the monopole parameter itself may trigger (terminate) the \textit{mode split} effect with the first (second) boundary condition. To make this point clear, we present the Maxwell QNMs in Fig.~\ref{Fig_monopole} by varying $8\pi\eta^2$, with fixed $N=0$, $\ell=1$ and $r_{+1}\approx4.237$ ($r_{+2}\approx2.505$) for the first (second) boundary condition. These BH radii are taken as the critical BH radii of $8\pi\eta^2=0.05$ for both boundary conditions. From Fig.~\ref{Fig_monopole}, it is shown evidently that, the mode with the first (second) boundary condition branches off (mergers) when $8\pi\eta^2>0.05$ since $r_{+1}$ ($r_{+2}$) we took is greater (smaller) than the critical BH radius.

\section{Discussion and Final Remarks}
\label{discussion}
In this paper, we have established that the Maxwell quasinormal spectrum may \textit{bifurcate} on asymptotically AdS BHs. To be specific, by taking Schwarzschild--AdS BHs both without and with a global monopole as examples, we unveil that when the BH radius is greater than the critical BH size, the real part of the Maxwell QNMs vanishes while the imaginary part branches off. This feature is coined as the \textit{mode split} effect, and is held for both boundary conditions. 

For the case of Schwarzschild--AdS BHs, we further explored the mode split effect by calculating the critical BH radius with respect to the overtone number $N$ and the angular momentum quantum number $\ell$. We observed that, for both boundary conditions, the critical BH radius increases as either $N$ or $\ell$ increases, and it scales exponentially with $N$ but linearly with $\ell$. 

An interesting aspect appeared for the monopole case is that, by increasing the monopole parameter, the critical BH radius decreases (increases) for the first (second) boundary condition. This property leads directly to a consequence that, by fixing a proper BH radius, the first (second) boundary condition may trigger (terminate) the mode split effect. 

To understand the universality and scrutinize the origin of the mode split effect, it is thus very interesting to extend this work to other asymptotically AdS BHs and the BH--mirror systems. Work along these directions is underway and we hope to report on them soon~\cite{wang}.

\bigskip

\noindent{\bf{\em Acknowledgements.}}
This work is supported by the National Natural Science Foundation of China under Grant Nos. 11705054, 11881240252, 11775076, 11875025, 12035005, and by the Hunan Provincial Natural Science Foundation of China under Grant Nos. 2018JJ3326 and 2016JJ1012. This study is also partially supported by the Innovative Experimental Project of College Students in Hunan Normal University (2018108).

\bibliographystyle{h-physrev4}
\bibliography{MaxwellGlobalAdS_PRD}


\end{document}